\documentclass[aps,prl,twocolumn,showpacs,superscriptaddress,groupedaddress,10pt]{revtex4-1}

\usepackage{graphicx}
\usepackage{dcolumn}
\usepackage{amsmath}
\usepackage{amssymb}
\usepackage[subrefformat=parens]{subcaption}

\usepackage{color}
\usepackage{ulem}

\newcommand{\asi}{$a$-Si}
\newcommand{\csi}{$c$-Si}
\newcommand{\Tfive}{T_5}
\newcommand{\Tthree}{T_3}
\newcommand{\Tfour}{T_{4\mathrm{a}}}

\begin{document}


\title{First-Principles Prediction of Densities of Amorphous Materials: The Case of Amorphous Silicon}
\author{Yoritaka Furukawa}
\author{Yu-ichiro Matsushita}
\affiliation{Department of Applied Physics, The University of Tokyo, Tokyo 113-8656, Japan}
\date{\today}

\begin{abstract}
A novel approach to predict the atomic densities of amorphous materials is explored on the basis of Car-Parrinello molecular dynamics (CPMD) in density functional theory.
Despite that determination of the atomic density of matter is crucial in understanding its physical properties, no such method has ever been proposed for amorphous materials until now.
In our approach, by assuming that the canonical distribution of amorphous materials is Gaussian distribution,
we generate multiple amorphous structures with several different volumes by CPMD simulations and average the total energies at each volume. The density is then determined to be the one that minimizes the averaged total energy.
In this study, this approach is implemented for amorphous silicon (\asi) to demonstrate its validity, and we have determined the density of {\asi} to be 4.1 \% lower and its bulk modulus to be 28 GPa smaller than those of the crystal, which are in good agreement with experiments.
We have also confirmed that generating samples through classical molecular dynamics simulations produces a comparable result and validates our assumption.
The findings suggest that the presented method is applicable to other amorphous systems, including those that lack experimental knowledge.
\end{abstract}
\maketitle

Determination of atomic structures is a prerequisite of any investigation of physical properties of condensed matter: Localization and delocalization of the electron wavefunction which is controlled by an atomic arrangement in a material are decisive in its physical properties.
The atomic density or the equilibrium volume of a material is also a crucial quantity that controls the behavior of its wavefunction.
Theoretical calculations based on the first principles of the quantum theory, e.g., density functional theory (DFT) \cite{hohenberg1964} with the Kohn-Sham scheme \cite{kohn1965}, contribute to the determination of the atomic density and the structure by minimizing the total energy of the system with respect to the atomic arrangement under a certain volume and then to the volumes.
However, most efforts in the past have focused on crystalline materials.
In amorphous materials, the determination of the atomic density and the structure is an inseparable issue since no one-to-one correspondence between the density and the structure is ensured, and thus has been rejecting theoretical approaches.
In this letter, we propose an approach based on Car-Parrinello Molecular Dynamics (CPMD) \cite{car1985} in DFT to determine the density of amorphous materials and confirm its validity for amorphous silicon (\asi).

Amorphous materials, which lack the long-range structural order but preserve the short-range order, provide a stage on which physics of disordered systems has been developed \cite{mott1971,brodsky1985,morigaki1999}.
From a technological viewpoint, {\asi} \cite{spear1975} and other amorphous materials, for instance, amorphous indium-gallium-zinc-oxide ($a$-IGZO) \cite{nomura2004}, are indispensable as flexible and superior materials for thin-film transistors.
In metal-oxide-semiconductor devices, which are ubiquitous in our life, insulating layers made up of amorphous $\mathrm{SiO_2}$ assure transistor actions in almost all electronic devices \cite{sze}.
Even a phase transition between the amorphous and the crystalline phases is utilized for memory devices \cite{wong2010}.
Not only semiconducting materials but amorphous metal alloys are also expected wider use for their mechanical strength and elasticity \cite{wang2009}.

Difficulty in determining the density of an amorphous material lies in a fact that the total energy of an amorphous material depends not only on its volume but also on its atomic arrangement: Even when the volume is fixed, each amorphous material has an entirely different atomic arrangement and consequently a different total energy.
This uncertainty of the total energy raises a fundamental question how one can define the density of an amorphous material.

Even though the total energies are not determined uniquely at each density,
it is expected that the total energies of amorphous structures are distributed near a certain value.
We here make an assumption that
at a fixed volume or density, the canonical distribution of an amorphous material forms a Gaussian distribution.
Under this assumption, whose validity will be discussed later in this paper, the amorphous material achieves a one-to-one correspondence between the density and the averaged total energy.
Then, one could determine the density of the amorphous material to be the one that minimizes the averaged energy.
The estimation of the ``true'' averaged energy is done by carefully generating multiple amorphous samples.

In our approach,
we generate amorphous samples by performing first-principles molecular dynamics simulations of which details are described below.
We then compute the total energy of each sample and compile the computed data in terms of the volume of the samples.
Then, the actual procedure is as follows:

\begin{enumerate}
    \item Generate $K$ amorphous samples for each of the volume in a set $\{ V_1, V_2, \dots, V_M \}$, thus having $K \times M$ different samples in total.
    \item Perform the DFT calculations to obtain total energies of all the samples.
    \item Average the total energies of the samples at each volume.
    \item Find a possible volume at which the averaged total energy achieves the minimum.
\end{enumerate}

In this work, we monitor $M=6$ volumes with $K=7$ samples for each volume, leading to the 42 samples in total.

In the approach above, the computational scheme to prepare amorphous structures is essential to ensure the reality of the obtained samples. There are several established ways, both empirical \cite{polk1971, weaire1974, biswas1987} and non-empirical \cite{car1988}, to generate amorphous samples. The non-empirical scheme based on the quantum theory has a significant advantage over the empirical scheme since it provides reliable interatomic forces based on the electronic-structure theory, thus being capable of describing chemical rebonding during the preparation of the amorphous structures. We here adopt CPMD based on DFT to describe atomic interactions.

Another important factor to ensure the validity of the simulation is the way to generate amorphous structures. Suitability of the melting procedure is assured by examining the atomic radial distribution in the liquid phase. Quenching, on the other hand, should be performed carefully. The CPMD simulations in the past to prepare amorphous structures used too fast quenching rates mainly due to the computational limitation \cite{stich1991}, and the prepared samples contained defects with unrealistically high concentration. This may cause severe artifacts in determining the density of amorphous materials.
We carefully do the quenching of the samples with the rate of 20 K/ps, which is the slowest ever implemented for {\asi}.

All calculations have been performed using our Real-Space-Density-Functional-Theory (RSDFT) code \cite{iwata2010,iwata2014,rsdft}, which employs the real-space scheme into the calculation of Kohn-Sham equations \cite{chelikowsky1994}. Generalized gradient approximation proposed by Perdew, Ernzerhof and Burke (PBE) \cite{perdew1996} is used for the exchange-correlation functional. It is known that PBE overestimates the lattice constant of crystalline silicon ({\csi}) by around 1 \%.

We have used a $3\times3\times3$ supercell model containing 54 Si atoms in its unitcell. As for the initial structures of CPMD simulations, we have prepared 6 {\csi} systems with different volumes. The volume of each system, $V$, is normalized to the calculated volume of {\csi}, $V_c$, obtained by our PBE calculation. The ratio $V/V_c$ takes 0.86, 0.91, 0.98, 1.03, 1.09, and 1.16, which is fixed throughout the simulation.

In the real-space scheme, grid points are introduced in the real space, and the wavefunction and the electron density are expanded on the mesh in the real space. Mesh spacing in the real-space grid is taken to be 0.48 {\AA}, corresponding to 40-Ryd cutoff energy in plane-wave-basis calculation. Integration over the Brillouin zone has been performed using the $\Gamma$-point. We have confirmed that these calculational conditions are sufficient to reproduce the experimental lattice constant of {\csi} within less than 1\% of the error. CPMD simulations have been done with 0.1 fs time step, and the temperature has been controlled by velocity scaling.

The melt-quench simulations to generate amorphous structures have been performed as follows. First, we have heated each system from 500 K to 1800 K with the constant heating rate 125 K/ps. Here, we have confirmed that all the systems have become liquids at the final step of heating. Subsequently, we have cooled the system with the constant cooling rate 20 K/ps until the temperature reaches 1000 K.
Finally, a static DFT calculation has been performed to relax the atomic configuration of the final step of the simulation, obtaining the stable geometry and its total energy.

To check whether each obtained structure is valid, we here examine the radial distribution $J(r)$, which is defined as
\begin{equation}
    J(r)=\frac{1}{N_s}\sum_{s=1}^{N_s}\left(\frac{1}{\rho N}\sum_{i \neq j}\delta(r-r_{ij}^s)\right)
    \label{eq:rdf}
\end{equation}
Here, $N_s$, $N$, $\rho$, and $r_{ij}^s$ are the total number of the simulation steps, the total number of atoms in the system, the density, and the distance between atom $i$ and $j$ at step $s$, respectively.
Delta function in Eq. (\ref{eq:rdf}) is modified into a form that is computationally treatable, namely,
\begin{equation}
    \delta(r)=
    \begin{cases}
        \displaystyle{\frac{1}{\Delta r}~~\left( -\frac{\Delta r}{2} \leq r < \frac{\Delta r}{2} \right),} \\
        0~~\left(\mathrm{other}\right)
    \end{cases}
\end{equation}
where $\Delta r$ is set to be a small number, 0.01 {\AA}.
To obtain $J(r)$ of each sample, a CPMD simulation has been additionally performed for 2 ps at 300 K, starting from the relaxed structure.

\begin{figure}[tb]
    \centering
    \includegraphics[width=50mm]{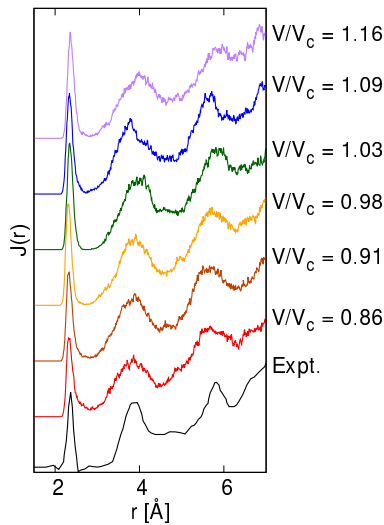}
    \caption{(Color online.) Radial distribution $J(r)$ of the obtained samples and an experimental one \cite{laaziri1999}. One sample has been chosen from at each volume ratio $V/V_c$.}
    \label{fig:rdf-asi}
\end{figure}

Figure \ref{fig:rdf-asi} shows $J(r)$ of one of the seven samples at each volume along with an experimental one \cite{laaziri1999}. We find that $J(r)$ is insensitive to the variations in the volume and that every $J(r)$ in the figure possesses following features. The first sharp peak is located around the bond length in {\csi}, 2.37 {\AA}, which suggests the existence of the short-range order in the structure. The subsequent peaks are much broader than the first peak, and this implies that the correlation of the atomic positions gets weaker as the distance increases, which should completely vanish at the long-distance limit. These observations are consistent with the experimentally obtained $J(r)$ \cite{laaziri1999}. Although Figure \ref{fig:rdf-asi} presents only one plot for each volume, we have confirmed that all the other plots follow the same characteristics. Thus, we can affirm that every sample is well amorphized from the crystal.

\begin{figure}[tb]
    \centering
    \includegraphics[width=65mm]{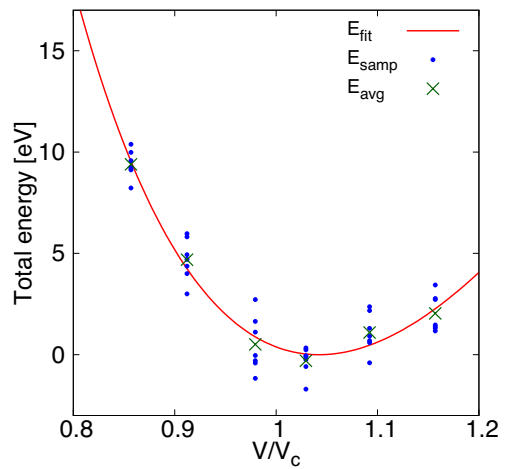}
    \caption{(Color online.) Calculated total energies of {\asi}. The total energy of each sample, $E_\mathrm{samp}$, the averaged total energy at each volume ratio, $E_\mathrm{avg}$, and the fitted energy, $E_\mathrm{fit}$, are represented by a blue dot, a green cross, and a red curve, respectively.
    All the energies are shifted so that the minimum of $E_\mathrm{fit}$, $E_\mathrm{min}$, becomes zero.}
    \label{fig:energy-asi}
\end{figure}

The total energy of each obtained sample, $E_\mathrm{samp}$, and the averaged total energy at each volume, $E_\mathrm{avg}$, are plotted in Figure \ref{fig:energy-asi}. We find that at each volume, $E_\mathrm{samp}$ is distributed within the range 3.9 eV, with the small variance 1.6 eV (see Table-S I in Supplemental Material for the energies of the samples, the average, and the variance at each volume \cite{supp}).
This finding suggests that the average $E_\mathrm{avg}$ is sufficiently meaningful for the evaluation of $V/V_c$.
To estimate the volume at which the structure achieves the minimum energy, Murnaghan's equation of state has been fitted to $E_\mathrm{avg}$. The fitted energy, $E_\mathrm{fit}$, is drawn in Figure \ref{fig:energy-asi} as a red curve.
From this fitting, we obtain the stable volume ratio $V_0/V_c$, where $E_\mathrm{fit}$ achieves its minimum $E_\mathrm{min}$, and the bulk modulus $B_0$. For comparison, we have additionally calculated the bulk modulus of {\csi}, $B_{0c}$. These values are presented in Table \ref{table:murnaghan},
along with the experimental ones \cite{custer1994,freund2003,szabadi1998,hopcroft2010}.
The remarkable finding in Table \ref{table:murnaghan} is not only that the determined $V_0/V_c$ and $B_0$ are comparable to the experimentally obtained ones, but that they also follow the same trend of the change from the crystal as in the experiment.
As shown in Table \ref{table:murnaghan}, {\asi} in the experiments has larger volume by 1.7 - 1.9 \%, or lower density by 1.7 - 1.9 \% \cite{custer1994}, and smaller $B_0$ by 2 - 62 GPa \cite{freund2003,szabadi1998} than those of {\csi}.
Similarly, {\asi} in our calculations has larger volume by 4.1 \%, or lower density by 4.1 \%, and smaller bulk modulus by 28.32 GPa than those of {\csi}.
This finding manifests that the proposed method produces physically reasonable results and is, therefore, applicable to {\asi}.

\begin{table}[tb]
    \begin{ruledtabular}
        \caption{Structural parameters obtained in our calculations and experiment:
        The volume ratio of {\asi} at stable state, $V_0/V_c$, the bulk modulus of {\asi}, $B_0$ [GPa], the bulk modulus of {\csi}, $B_{0c}$ [GPa], and the difference of the bulk modulus between the crystalline and the amorphous phases, $B_{0c} - B_0$ [GPa].}
        \begin{tabular}{ccc}
                           & This work & Expt. \\
            \hline
            $V_0/V_c$      & 1.042     &  1.017 - 1.019   \cite{custer1994}   \\
            $B_0$          & 61.27     &  36 - 60 \cite{freund2003};  86 - 95 \cite{szabadi1998} \\
            $B_{0c}$       & 89.59     &  97.6            \cite{hopcroft2010} \\
            $B_{0c}-B_0$   & 28.32     &  2 - 62                              \\
        \end{tabular}
    \label{table:murnaghan}
    \end{ruledtabular}
\end{table}

Besides, we have clarified the relationship between the atomic density and the number of defects found in the samples. In addition to the shape of $J(r)$, defect concentration is also an important property of {\asi} because defects contribute to the degradation of the carrier mobility caused by dangling bonds and floating bonds, the latter of which are the state that comprises a five-coordination of silicon
\cite{pantelides1986,biswas1989}.
Defects in {\asi} can be classified into the three distinct ones: three-fold ($\Tthree$), five-fold ($\Tfive$) and anomalous four-fold ($\Tfour$) defects, whose bond angles are heavily distorted from the tetrahedral bonds while maintaining chemical bonds with the four neighboring atoms.
For realisticity, the defect densities in the computationally generated structures are expected to be close to what is found in the experiment $\sim 0.1$ \% \cite{brodsky1979}, which indicates that the number of defects in our 54-atom system should be close to 0 or 1 per supercell.
The numbers of the three kinds of defects found in our samples are presented in Figure \ref{fig:defects-asi} (see Table-S II in Supplemental Material for the specific numbers of the defects \cite{supp}).
To calculate the coordination number of each atom, we have set the threshold of the bond length as 2.8 {\AA}.

In Figure \ref{fig:defects-asi}, the $\Tfive$ defects are dominant at every volume. The number of {$\Tfive$} defects becomes drastically larger at the smaller volumes, reaching up to $\simeq$ 23, while at $V/V_c > 1$ it is less than 3.
In contrast, the number of $\Tthree$ tends to increase along with the volume, its maximum being less than 2. The number of $\Tfour$ is comparable to that of $\Tthree$ defects, having its range within 0 and 2.
We find that at $V/V_c=1.03$, which is the closest to the predicted volume ratio, 1.042, the number of $\Tfive$ and $\Tfour$ defects takes the minimum, and the number of every defect is less than 2, which is certainly close to that in experiments.
Thus, our approach has successfully reproduced these structural properties as well as the atomic density.

\begin{figure}[tb]
    \centering
    \includegraphics[width=70mm]{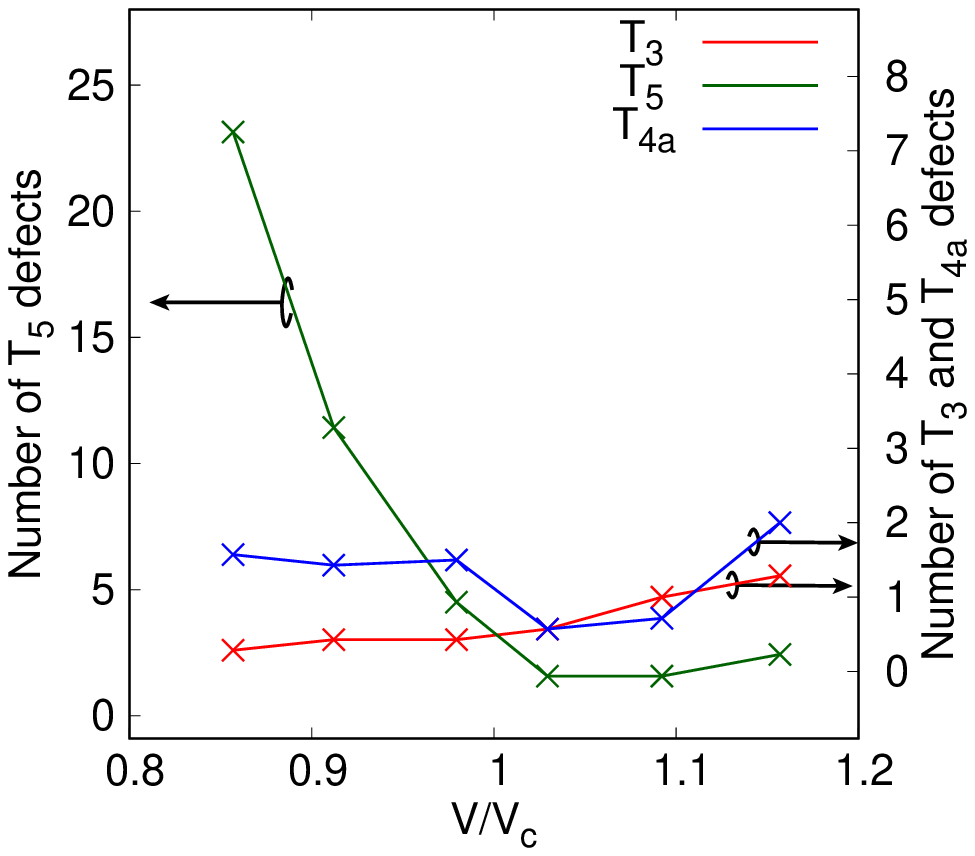}
    \caption{(Color online.) Averaged number of each {$\Tfive$}, {$\Tthree$}, and {$\Tfour$} defects per supercell.}
    \label{fig:defects-asi}
\end{figure}

Also, we point out that we have checked the validity of employing 54-atom systems for this study, by confirming that the total energy per cell and the defect densities are consistent with those in a larger system. The details are described in Supplemental Material.

Finally, we briefly report that we have examined the validity of our method when employing a classical molecular dynamics (MD) scheme for the generation of the samples instead of the CPMD scheme (See Supplemental Material for the details of the calculations).
Using the Tersoff potential \cite{tersoff1988}, we have generated 30 different samples for each volume, taking advantage of its low computational cost.
The stable volume has been determined to be -0.8 \% smaller than that of crystal, which is close to the experimental difference (1.8 \% of increase), and the bulk modulus has also been found to the comparable to the experiments.
We note, however, that the total energy at the stable volume is larger by 3 eV/cell than that has been obtained from the CPMD simulations, which indicates that the structures generated by CPMD simulations are energetically preferable.

\begin{figure}[tb]
    \centering
    \includegraphics[width=65mm]{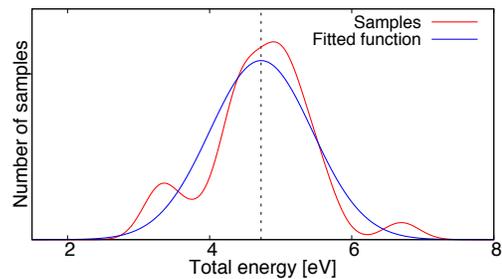}
    \caption{(Color online.)
    The distribution of the total energy of the samples per supercell at $V/V_c=0.86$ obtained from the MD simulations.
    The red curve is the sum of all the 30 samples, each of which has been smeared.
    The blue curve is the Gaussian function fitted to the obtained data.
    The dashed line indicates the averaged energy.
    }
    \label{fig:gaussian}
\end{figure}

Another important finding with the MD scheme is that the distribution of the total energies of the samples is well fitted to the Gaussian distribution.
Figure \ref{fig:gaussian} explicitly shows the distribution at $V/V_c=0.86$ with the fitted Gaussian curve, each drawn in red and blue.
This certainly supports our underlying assumption, thus strengthening the validity of our method.

To conclude, we have proposed a first-principles method to predict densities of amorphous materials for the first time and applied it to {\asi} for its demonstration.
We have introduced an assumption that the canonical distribution of amorphous materials is Gaussian distribution.
Under this assumption, we have generated multiple {\asi} samples with several volumes, whose atomic configurations are completely different from each other, and calculated the average of their total energies at each volume.
The stable volume, and hence the density, has been determined to be the one that minimizes the averaged energy.
The determined density of {\asi} has been lower than that of crystal by 4.1 \%, which is comparable to the experimental one, 1.8 \%.
The results are physically meaningful because not only the density but the bulk modulus also decrease from the crystal, by 28 GPa, in the same manner as in the experiment.
In addition, we have found that sampling through classical MD simulations also achieves comparable results and that the distribution of the total energies of the samples supports our assumption.
This consequence indicates that the proposed method is applicable to the computational studies of other types of amorphous materials, such as those whose experimental properties are yet to be identified.

\begin{acknowledgements}
We appreciate Professor Atsushi Oshiyama for fruitful discussions. This work has been supported in part by Ministry of Education, Culture, Sports, Science and Technology. Computations were performed mainly at the Supercomputer Center at the Institute for Solid State Physics, The University of Tokyo, The Research Center for Computational Science, National Institutes of Natural Sciences, and the Center for Computational Science, University of Tsukuba. This research partly used computational resources of the K computer provided by the RIKEN Advanced Institute for Computational Science through the HPCI System Research project (Project ID:hp160265). This work was supported by JSPS Grant-in-Aid for Young Scientists (B) Grant Number 16K18075.
\end{acknowledgements}

\end{document}